\documentclass[aps,prd,nofootinbib,amsmath,amssymb,superscriptaddress,preprintnumbers,twocolumn,10pt]{revtex4}

\usepackage{graphics,epsfig,subfigure}
\usepackage{ulem}
\usepackage{color}
\usepackage{url}
\usepackage{float}
\usepackage{dcolumn}
\usepackage{bm}
\usepackage{amssymb}
\usepackage{latexsym}
\usepackage{booktabs}
\usepackage{amsmath}
\allowdisplaybreaks[4]
\usepackage{multirow}
\usepackage[colorlinks=true, linkcolor=red, citecolor=blue]{hyperref}

\newcommand{\be}{\begin{equation}}
\newcommand{\ee}{\end{equation}}
\newcommand{\bq}{\begin{eqnarray}}
\newcommand{\eq}{\end{eqnarray}}

\begin{document}
\newcommand{\PR}[1]{\ensuremath{\left[#1\right]}} 
\newcommand{\PC}[1]{\ensuremath{\left(#1\right)}} 
\newcommand{\PX}[1]{\ensuremath{\left\lbrace#1\right\rbrace}} 
\newcommand{\BR}[1]{\ensuremath{\left\langle#1\right\vert}} 
\newcommand{\KT}[1]{\ensuremath{\left\vert#1\right\rangle}} 
\newcommand{\MD}[1]{\ensuremath{\left\vert#1\right\vert}} 

\title{Probing a polymerized black hole with the frequency shifts of photons}
\author{Qi-Ming Fu}\thanks{fuqiming@snut.edu.cn}
\affiliation{Key Laboratory of Cosmology and Astrophysics (Liaoning Province), College of Sciences, Northeastern University, Shenyang 110819, China}
\affiliation{Institute of Physics, Shaanxi University of Technology, Hanzhong 723000, China}

\author{Xin Zhang}\thanks{Corresponding author. \\zhangxin@mail.neu.edu.cn}
\affiliation{Key Laboratory of Cosmology and Astrophysics (Liaoning Province), College of Sciences, Northeastern University, Shenyang 110819, China}
\affiliation{Key Laboratory of Data Analytics and Optimization for Smart Industry (Ministry of Education), Northeastern University, Shenyang 110819, China}
\affiliation{National Frontiers Science Center for Industrial Intelligence and Systems Optimization, Northeastern University, Shenyang 110819, China}

\begin{abstract}
As is well-known, black hole plays an important role in probing the quantum effects of gravity in the strong-field regime.
In this paper, we focus on a polymerized black hole and investigate the quantum effects on the redshift, blueshift and gravitational redshift of photons emitted by massive particles revolving around this black hole. With a general relativistic formalism, we obtain two elegant and concise expressions for the mass of the black hole and the quantum parameter in terms of few direct observables, 
and find that all the frequency shifts of photons decrease with the quantum parameter. 
{By expanding the total frequency shift for large orbit of the emitter, low peculiar redshift, and small azimuthal angle, we find that the leading contributions to the frequency shift are originated from the kinematic shift and peculiar shift, and the quantum effects make a negligible contribution.}
In addition, we study the effects of the plasma on the redshift/blueshift for the first time, and conclude that the presence of plasma results in the decrease of the redshift/blueshift.
\end{abstract}

\maketitle

\section{Introduction}

Although general relativity has been proven to be very successful in modern physics, it may not be the final theory of gravity because of its prediction of spacetime singularity inside a black hole or at the start of the Universe, which indicates that the theory invalidates there. It is widely believed that the singularity can be avoided by incorporating the quantum effects. Many efforts have been devoted to address these issues by resorting to the quantum theory of gravity. One of the most successful attempts is to apply an effective approach developed in loop quantum gravity to the big-bang singularity in cosmological scenarios, where the singularity is replaced by a quantum bounce bridging a contracting branch with an expanding one \cite{Ashtekar084003,Ashtekar213001}. The key point is the phase space quantization usually called polymerization \cite{Corichi044016}. This quantization procedure introduces a natural cutoff called the polymer scale, where the quantum effects become relevant close to this scale. 

In recent years, similar technique has been extended to the context of black holes, and a lot of polymerized black hole models have been constructed \cite{Ashtekar391,Modesto5587,Campiglia3649,Boehmer104030,Chiou064040,Corichi055006,Olmedo225011,Bodendorfer187001,Bodendorfer195015,Kelly106024,Faraoni1264,Gan124030,Bojowald046006,Sartini066014,Bodendorfer136390,Bodendorfer095002}. Most of these models focused on the spherically symmetric cases. The common feature of these types of models is that the singularity inside a black hole is replaced by a spacelike transition hypersurface, which smoothly connects  a trapped region (the black hole region) to an antitrapped region (the white hole region), and the spacetime is regular everywhere. Besides, the resulting spacetime with the quantum corrections can still be described by an explicit metric. Thus, one can investigate such nonsingular models by exploiting the already well-developed mechanism for the semiclassical black holes.

In the last couple of years, various aspects of the polymerized black holes have been studied in the literature. For example, the role of Dirac observables in polymerized black hole models was investigated in Ref.~\cite{Bodendorfer095002}, and the interior structure, quasinormal spectra of various perturbation modes were studied in Ref.~\cite{Bouhmadi066}. The thermodynamic properties of the  polymerized black holes were analyzed in Ref.~\cite{Mele011}, and the effects of the quantum corrections on the relevant thermodynamic quantities also were considered. In addition, the Hawking radiation emitted by polymerized black holes for the fields with different spins were investigated in Refs.~\cite{Arbey104010,Arbey084016}, and the authors also investigated the primordial black hole evaporation signals with the Hawking radiation rates. Recently, the strong gravitational lensing of the polymerized black holes was investigated in Ref.~\cite{Fu064020}, and the solar system experiments of these black holes were considered in Ref.~\cite{Liu084068}, from which the constraints on the polymer scale were derived as well. Other phenomenological investigations and observational constraints on the polymerized black holes were explored in Refs.~\cite{Brahma181301,Achour041,Achour124041,Achour020,Blanchette084038,Munch046019,Liu02861,Walia02106,Yang066} and references therein. In this paper, we mainly concentrate on {a particular polymerized black hole introduced in Refs.~\cite{Bodendorfer136390,Bodendorfer095002}} and investigate the frequency shifts of the photons emitted by massive objects orbiting this black hole.

On the other hand, the authors of Ref.~\cite{Herrera045024} introduced a general relativistic method that can express the mass and spin of a Kerr black hole in terms of direct observables, i.e., the radius of the massive particles orbiting the black hole and the frequency shifts experienced by the photons emitted by these revolving particles. Subsequently, this general relativistic mechanism was applied to other black holes, such as, Myers-Perry black hole \cite{Sharif404}, Kerr-Newman-(anti) de Sitter black hole \cite{Kraniotis147}, and Plebanski-Demianski black hole \cite{Ujjal213}. Besides, the redshift and blueshift of photons emitted by timelike geodesic particles near  compact objects, such as, boson stars, as well as the Schwarzschild and Reissner-Nordstr$\ddot{\text{o}}$m black holes, were investigated in Ref.~\cite{Becerril124024}. In Refs.~\cite{Lopez64,Becerril084054}, a similar formalism was used to obtain the relations between the parameters describing a class of regular black holes and the redshift observables. Other investigations on the frequency shifts of photons can be found in Refs.~\cite{Rashmi025003,Sheoran124049,Komarov132,Lopez55,Herrera198,Banerjee124037}.

Although the polymerized black hole has been extensively studied in the literature, the frequency shifts of photons emitted by massive particles orbiting this black hole have not been analyzed yet. With the general relativistic formalism, one can obtain closed formulas for the mass and the quantum parameter of the polymerized black hole in terms of the frequency shifts of the emitted photons. Thus, this mechanism not only provides us with valuable information about the black hole with few direct observables, but also can be used as a powerful tool to probe the quantum effects. On the other hand, most astrophysical black holes in nature are surrounded by a plasma medium and the frequency of photons will be influenced by the plasma \cite{Rogers17,Perlick104031,Atamurotov084005}. 
Thus, it is necessary to study the polymerized black hole with the general relativistic formalism and it is also of interest to know what the effects are of the plasma on the frequency shifts.

{In this paper, we focus on exploring the quantum effects through the frequency shifts of photons for a particular polymerized black hole model, and give a brief analysis about the effects of the plasma on the frequency shifts.} This paper is organized as follows: In Sec.~\ref{BHs}, we first give a brief review of the polymerized black hole and then derive the equations of motion for the orbiting massive particles and the photons emitted by these particles. In Sec.~\ref{RB}, we investigate the frequency shifts of the photons and derive the closed formulas for the mass and quantum parameter of the black hole. The effects of the plasma on the frequency shifts are analyzed in Sec.~\ref{RBp}. Section~\ref{con} contains the conclusion.

\section{Particles and photons in a polymerized black hole spacetime}~\label{BHs}

The static and spherically-symmetric metric modified by the effective loop quantum gravity can be expressed as \cite{Bodendorfer136390,Bodendorfer095002,Brahma181301},
\begin{eqnarray}
ds^2=-\mathcal{A}(y)d\tau^2+\frac{dy^2}{\mathcal{A}(y)}+\mathcal{C}^2(y)d\Omega_2^2, ~\label{lqgbh}
\end{eqnarray}
where the radial coordinate $y\in (-\infty,\infty)$, and the metric functions are given by
\begin{eqnarray}
\mathcal{A}(y)&=&\left(1-\frac{1}{\sqrt{2A_{\lambda}(1+y^2)}}\right)\frac{1+y^2}{\mathcal{C}^2(y)}, \\
\mathcal{C}^2(y)&=&\frac{A_{\lambda}}{\sqrt{1+y^2}}\frac{M_B^2(y+\sqrt{1+y^2})^6+M_W^2}{(y+\sqrt{1+y^2})^3}.
\end{eqnarray}
Here $M_B$ and $M_W$ represent the masses of an asymptotically Schwarzschild black hole and a white hole, respectively \cite{Brahma181301}, and $A_{\lambda}$ is a dimensionless and non-negative parameter defined by $A_{\lambda}\equiv (\lambda_k/(M_B M_W))^{2/3}/2$, where $\lambda_k$ is a quantum parameter originated from holonomy modifications {and should be thought of being of Planckian order \cite{Bodendorfer136390,Bodendorfer095002}. Besides, according to its definition, the parameter $A_{\lambda}$ is extremely small for astrophysical black holes due to the large suppression of the black hole mass.}  In this paper, we focus on the case of $M_B=M_W=M$ corresponding to a symmetric bounce, i.e., the spacetime is symmetric under $y\rightarrow -y$. Without loss of generality, we focus on the positive branch with $y\geqslant 0$. By redefining two new coordinates $r\equiv\sqrt{8A_{\lambda}}My$ and $t\equiv\tau/(\sqrt{8A_{\lambda}}M)$, the metric (\ref{lqgbh}) can be rexpressed as \cite{Brahma181301,Fu064020}
\begin{eqnarray}
ds^2=-A(r)dt^2+B(r)dr^2+C(r)(d\theta^2+\sin^2\theta d\phi^2), ~\label{metric1}
\end{eqnarray}
where
\begin{eqnarray}
A(r)\!\!&=&\!\!\frac{1}{B(r)}\!=\!\frac{\sqrt{8A_{\lambda}M^2+r^2}\big(\sqrt{8A_{\lambda}M^2+r^2}-2M\big)}{2A_{\lambda}M^2+r^2},~\label{metricA} ~~~\\
C(r)\!\!&=&\!\!2A_{\lambda}M^2+r^2~\label{metricC}.
\end{eqnarray}
The event horizon of this black hole is determined by $A(r)=0$, which admits the solution $r=2 M \sqrt{1-2 A_{\lambda }}$. Hence the permissible range for the quantum parameter $A_{\lambda}$ in this black hole spacetime is $0\leqslant A_{\lambda} \leqslant \frac{1}{2}$.

The Lagrangian for a massive particle moving around  this black hole is
\begin{eqnarray}
2\mathcal{L}=-A(r)\dot{t}^2+B(r)\dot{r}^2+C(r)(\dot{\theta}^2+\sin^2\theta\dot{\phi}^2), ~\label{la}
\end{eqnarray}
where $\dot{x}^{\mu}=U^{\mu}\equiv\frac{dx^{\mu}}{d\tau}$, $U^{\mu}$ denotes the four velocity of the massive particle, and $\tau$ is an affine parameter along the geodesic. Since the Lagrangian only depends on $r$ and $\theta$, there appear two conserved quantities along the geodesic,
\begin{eqnarray}
p_t&=&\frac{\partial\mathcal{L}}{\partial\dot{t}}=-A(r)\dot{t}=-A(r)U^t=-E, ~\label{pt} \\
p_{\phi}&=&\frac{\partial\mathcal{L}}{\partial\dot{\phi}}=C(r)\sin^2\theta\dot{\phi}=C(r)\sin^2\theta U^{\phi}=L, ~\label{pphi}
\end{eqnarray}
which correspond to the energy and angular momentum of the massive particle with unit rest mass, respectively. From the normalized condition $U_{\mu}U^{\mu}=-1$, the geodesic equation for the massive particle at the equatorial plane ($U^{\theta}=0$) can be derived as
\begin{eqnarray}
(U^r)^2+V(r)=0,
\end{eqnarray}
with the effective potential
\begin{eqnarray}
V(r)=\frac{1}{B}\left(1-\frac{E^2}{A}+\frac{L^2}{C}\right).
\end{eqnarray}

For circular orbits, i.e., $U^r=0$, the effective potential must have an extremum indicating two conditions $V(r)=0$ and $V'(r)=0$, where the prime stands for the derivative with respect to $r$. From these two conditions, one can obtain
\begin{eqnarray}
E^2&=&\frac{A^2 C'}{A C'-C A'}, ~\label{E2} \\
L^2&=&\frac{C^2 A'}{A C'-C A'}. ~\label{L2}
\end{eqnarray}
Besides, for the circular orbits to be stable, the following stable condition should also be satisfied:
\begin{eqnarray}~\label{Veffdd}
V''\!\!=\!\!\frac{C' \!\left(2 A A' C'\!+\!C \left(A A''\!-\!2 A'^2\right)\right)\!-\!A C A' C''}{A B C \left(A C'-C A'\right)}>0.
\end{eqnarray}
Inserting Eqs.~(\ref{E2}) and (\ref{L2}) into Eqs.~(\ref{pt}) and (\ref{pphi}), the four-velocity for the circular and equatorial orbiting massive particles can be calculated as
\begin{eqnarray}
U^t&=&\sqrt{\frac{C'}{A C'-C A'}}, ~\label{Ut} \\
U^{\phi}&=&\sqrt{\frac{A'}{A C'-C A'}}. ~\label{Uphi}
\end{eqnarray}

On the other hand, the energy and angular momentum for the photons moving along null geodesics can be obtained in a similar way,
\begin{eqnarray}
E_{\gamma}=A k^t,  \quad L_{\gamma}=C k^{\phi}, ~\label{ELgamma}
\end{eqnarray}
with $k^{\mu}=(k^t, k^r, k^{\theta}, k^{\phi})$ the four momentum of the photons, which satisfies $k^{\mu}k_{\mu}=0$. Then, the equation of motion for the photons at the equatorial plane is
\begin{eqnarray}
-A(k^t)^2+B(k^r)^2+C(k^{\phi})^2=0. ~\label{eomphoton}
\end{eqnarray}
After inserting Eq.~(\ref{ELgamma}) into the above equation, the radial component of the four momentum can be solved as
\begin{eqnarray}~\label{kr}
(k^r)^2=\frac{-A L_{\gamma}^2+C E_{\gamma}^2}{ABC}.
\end{eqnarray}

Introducing an auxiliary bidimensional and geometrical vector defined by $\kappa^2\equiv (k^r)^2+C(k^{\phi})^2$, where the components $k^r$ and $k^{\phi}$ satisfy the following decomposition \cite{Banerjee124037}:
\begin{eqnarray}
k^r&=&\kappa \cos\phi, ~\label{kkr} \\
\sqrt{C} k^{\phi}&=&\kappa \sin\phi,
\end{eqnarray}
with $0\leqslant \phi \leqslant 2\pi$, the auxiliary vector can be explicitly expressed as
\begin{eqnarray}~\label{kappa1}
\kappa^2=\frac{-A L_{\gamma}^2+C E_{\gamma}^2}{ABC}+C\left(\frac{L_{\gamma}}{C}\right)^2.
\end{eqnarray}
Besides, after inserting Eq.~(\ref{kkr}) into Eq.~(\ref{kr}), another expression for the auxiliary vector can be obtained as
\begin{eqnarray}~\label{kappa2}
\kappa^2=\frac{-A L_{\gamma}^2+C E_{\gamma}^2}{ABC\cos^2\phi}.
\end{eqnarray}

Combining these two expressions (\ref{kappa1}) and (\ref{kappa2}) for the auxiliary vector together, one can obtain an equation for the impact parameter $b\equiv \frac{E_{\gamma}}{L_{\gamma}}$,
\begin{eqnarray}
b^2A(\sin^2\phi+B\cos^2\phi)-C\sin^2\phi =0,
\end{eqnarray}
which admits the following solution
\begin{eqnarray}
b=-\frac{\sqrt{C}\sin\phi}{\sqrt{A(\sin^2\phi+B\cos^2\phi)}}. ~\label{b}
\end{eqnarray}
The maximum or minimum of the impact parameter is reached when $k^r=0$ corresponding to $\phi=-\pi/2$ or $\pi/2$, respectively.

\section{Redshift/Blueshift of emitted photons}~\label{RB}

The frequency shift of a photon emitted by a massive particle moving around a black hole and then received by a detector can be defined as \cite{Herrera045024}
\begin{eqnarray}~\label{zcomplete}
1+z=\frac{\omega_{\text{e}}}{\omega_{\text{d}}}=\frac{-k_{\mu}U^{\mu}|_{\text{e}}}{-k_{\mu}U^{\mu}|_{\text{d}}},
\end{eqnarray}
where the lower indexes $\text{e}$ and $\text{d}$ represent the emitter and detector, respectively. For the case of a circular and equatorial orbiting massive particle, i.e., $U^r=U^{\theta}=0$, Eq.~(\ref{zcomplete}) can be reduced to
\begin{eqnarray}~\label{zsimplify}
1+z=\frac{(E_{\gamma}U^t-L_{\gamma}U^{\phi})|_{\text{e}}}{(E_{\gamma}U^t-L_{\gamma}U^{\phi})|_{\text{d}}}=\frac{U^t_{\text{e}}-b_{\text{e}}U^{\phi}_{\text{e}}}{U^t_{\text{d}}-b_{\text{d}}U^{\phi}_{\text{d}}}.
\end{eqnarray}

There are mainly two contributions to the frequency shift defined above, i.e., the gravitational effect and the kinematic effect or Doppler effect. The impact parameter for a radially emitted photon is usually called the central impact parameter $b_{\text{c}}$, which vanishes for a spherically-symmetric and static black hole, and the only contribution to the frequency shift of this photon is the gravitational effect. Thus, the gravitational redshift can be conveniently defined as \cite{Becerril124024}
\begin{eqnarray}~\label{zg}
1+z_{\text{g}}=\frac{U^t_{\text{e}}-b_{\text{c}}U^{\phi}_{\text{e}}}{U^t_{\text{d}}-b_{\text{c}}U^{\phi}_{\text{d}}}=\frac{U^t_{\text{e}}}{U^t_{\text{d}}}.
\end{eqnarray}
Then, the kinematic redshift $z_{\text{kin}}$ originated from the kinematic effect of the emitter can be defined by subtracting the gravitational redshift $z_{\text{g}}$ from the frequency shift $z$ \cite{Becerril124024},
\begin{eqnarray}~\label{zkin}
z_{\text{kin}}\equiv z-z_{\text{g}}=\frac{U^t_{\text{e}}-b_{\text{e}}U^{\phi}_{\text{e}}}{U^t_{\text{d}}-b_{\text{d}}U^{\phi}_{\text{d}}}-\frac{U^t_{\text{e}}}{U^t_{\text{d}}}.
\end{eqnarray}

Besides, if there is a relative motion of the black hole with respect to a distant observer or detector, there should be an additional correction to the frequency shifts of photons coming from the special relativistic boost, which is defined by \cite{Banerjee124037}
\begin{eqnarray}
1+z_{\text{boost}}=\gamma(1+\beta), \quad \gamma=(1-\beta^2)^{-1/2}, \quad \beta=\frac{v_0}{c},~~
\end{eqnarray}
where $v_0=z_0 c$ represents the radial peculiar velocity of the black hole with respect to the observer, and $z_0$ is called the peculiar redshift which codifies the motion of the black hole receding (positive $z_0$) or approaching (negative $z_0$) with respect to the observer. Then, by considering Eqs.~(\ref{Ut}), (\ref{Uphi}) and (\ref{b}), the total frequency shift can be expressed as \cite{Banerjee124037}
\begin{eqnarray}~\label{ztot}
z_{\text{tot}}&=&(1+z)(1+z_{\text{boost}})-1 \nonumber\\
&=&(1+z_{\text{kin}}+z_{\text{g}})\gamma(1+\beta)-1 \nonumber\\
&=&\sqrt{\frac{1+z_0}{1-z_0}}\bigg(\sqrt{\frac{C'(r_\text{e})}{A(r_\text{e}) C'(r_\text{e})-C(r_\text{e}) A'(r_\text{e})}} \nonumber\\
&+&\frac{\sin\phi \sqrt{\frac{C(r_\text{e})A'(r_\text{e})}{A(r_\text{e}) C'(r_\text{e})-C(r_\text{e}) A'(r_\text{e})}}}{\sqrt{A(r_\text{e})(\sin^2\phi+B(r_\text{e})\cos^2\phi)}}\bigg)-1,
\end{eqnarray}
where $r_\text{e}$ is the orbital radius of the revolving massive particles, and the static observer is assumed to be located at infinity with the four velocity $U^{\mu}|_{\text{d}}=(1,0,0,0)$.

{By substituting the metric functions (\ref{metricA}) and (\ref{metricC}) into Eq.~(\ref{ztot}), one can obtain the total frequency shift for this particular polymerized black hole, which is not given here due to its length. However, for large orbit of the emitter $M\ll r_{\text{e}}$, low peculiar redshift ($z_0\ll 1$), and small azimuthal angle $\phi\approx 0$, the total frequency shift can be expanded as
\begin{eqnarray}
z_{\text{tot}}&\approx&\frac{3}{2}\tilde{M}+z_0+\phi\tilde{M}^{1/2}+\frac{27\tilde{M}^2}{8}+\frac{3\tilde{M}z_0}{2}+\frac{1}{2}z_0^2 \nonumber\\
&+&\frac{3}{2}\tilde{M}^{3/2}\phi+\tilde{M}^{1/2}z_0\phi-6A_{\lambda}\tilde{M}^2,
\end{eqnarray}
with $\tilde{M}\equiv M/r_{\text{e}}$. Obviously, the first two terms are the leading contribution to the frequency shift. The first term is originated from the so-called Doppler or kinematic shift, i.e., the motion of the emitter revoloving around the black hole. The second term is originated from the peculiar motion, i.e., the emitter and the black hole as a whole entity approaching or receding from us. The third term is the subleading contribution to the frequency shift which is also originated from the kinematic shift. The fifth and eighth terms in this expansion are originated from the combination of the kinematic shift and peculiar redshift.
The parameter $A_{\lambda}$ first appears in the ninth term of this series expansion, which encodes the effects of the quantum corrections and indicates a very subtle contribution to the total frequency shift. In addition, $A_{\lambda}$ is extremely small for astrophysical black holes also signifying that the quantum effects on the frequency shift can be negligible.
}

Before proceeding, we should note that there exist three special azimuthal angles, i.e., $\phi=0, \pm\frac{\pi}{2}$. Here $\phi=0$ is assumed to be the opposite direction of the line of sight of the observer, and the observer is on the right-hand side of the black hole. Obviously, the impact parameter vanishes in this case and describes a photon emitted by an anticlockwise orbiting particle when moving at the location with the azimuthal angle $\phi=0$. At this moment, the three velocity of the orbiting particle is perpendicular with respect to the line of sight of the distant observer, and the only contribution to the frequency shift of photons is the gravitational effect, which is defined by $z_{\text{tot}_3}\equiv z_{\text{tot}}|_{\phi\rightarrow 0}$. Then, Eq.~(\ref{b}) shows that $\phi=\pm\frac{\pi}{2}$ corresponds to the minimum and maximum of the impact parameter, respectively. The minimal impact parameter describes the photon emitted by the anticlockwise orbiting particle when arriving at the location with the azimuthal angle $\phi=\frac{\pi}{2}$. At this moment, the initial three velocity of the photon is exactly opposite to the instantaneous orbital velocity of the particle. Thus, the redshift of the photon reaches the maximum defined by $z_{\text{tot}_1}\equiv z_{\text{tot}}|_{\phi\rightarrow \pi/2}$. On the other hand, the maximal impact parameter describes the photon emitted by the revolving particle when reaching at the location with the azimuthal angle $\phi=-\frac{\pi}{2}$ and the initial three-velocity of the photon is the same as the instantaneous orbital velocity of the particle, which corresponds to the case of the maximal blueshift of the photon defined by $z_{\text{tot}_2}\equiv z_{\text{tot}}|_{\phi\rightarrow -\pi/2}$.

Then, by defining the following variables,
\begin{subequations} ~\label{RSTd}
\begin{eqnarray}
R&\equiv& 1+z_{\text{tot}_1}, \\
S&\equiv& 1+z_{\text{tot}_2}, \\
T&\equiv& \left(1+z_{\text{tot}_3}\right)^2,
\end{eqnarray}
\end{subequations}
one can obtain two elegant closed formulas for the mass of the black hole and the quantum parameter in terms of these variables by inverting the relations (\ref{RSTd}):
\begin{eqnarray}
M\!\!&=&\!\!r_\text{e} \left(1+\frac{\zeta }{T }-\frac{2 \zeta }{R S}\right) \sqrt{\frac{3 \zeta  T }{15 \zeta  T -4 R S (2 \zeta +T )}},~\label{Mexp}\\
A_{\lambda}\!\!&=&\!\!\frac{T  R^2 S^2 (R S (2 \zeta +T )-3 \zeta  T )}{6 \zeta  (R S (\zeta +T )-2 \zeta  T )^2},
\end{eqnarray}
with $\zeta\equiv \frac{1+z_0}{1-z_0}$.
{The function $M$ given by Eq.~(\ref{Mexp}) and the radius of the emitter $r_{\text{e}}$ are expressed in geometrized units ($G=c=1$) and scaled by  $p M_{\odot}$, where $M_{\odot}$ is the solar mass and $p$ is an arbitrary factor of proportionality, such as, $p=4.3\times 10^6$ for SgrA*.}
These two formulas tell us that the mass of the black hole and the quantum parameter can be completely and independently given by few direct observables, i.e., the redshift $z_{\text{tot}_1}$, blueshift $z_{\text{tot}_2}$, and gravitational redshift $z_{\text{tot}_3}$ of photons, as well as the orbital radius $r_{\text{e}}$ of the massive particle. It should be noted that the peculiar redshift $z_0$ is not a measurable quantity, but can be statistically estimated with the help of the direct observables and Eq.~(\ref{ztot}). Obviously, this general relativistic formalism described above can be used to predict the property of the compact object and probe the quantum effects precisely. To be specific, the compact object is a Schwarzschild black hole for $A_{\lambda}=0$, a regular black hole for $0<A_{\lambda}<\frac{1}{2}$, and a wormhole for $A_{\lambda}=\frac{1}{2}$. One can see Ref.~\cite{Fu064020} for more details.

\begin{figure}[htb]
\begin{center}
\subfigure[]  {\label{A}
\includegraphics[width=7cm]{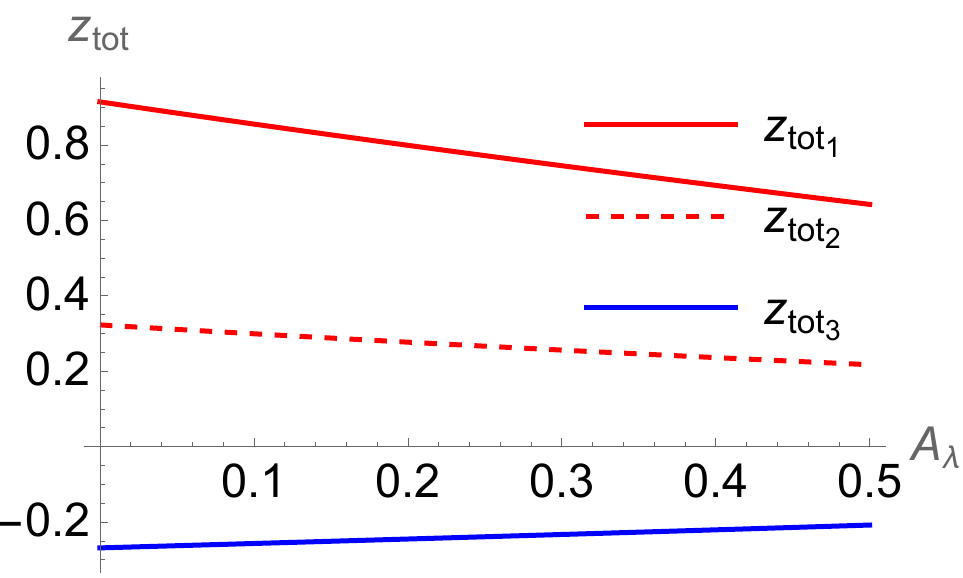}}
\subfigure[]  {\label{re}
\includegraphics[width=7cm]{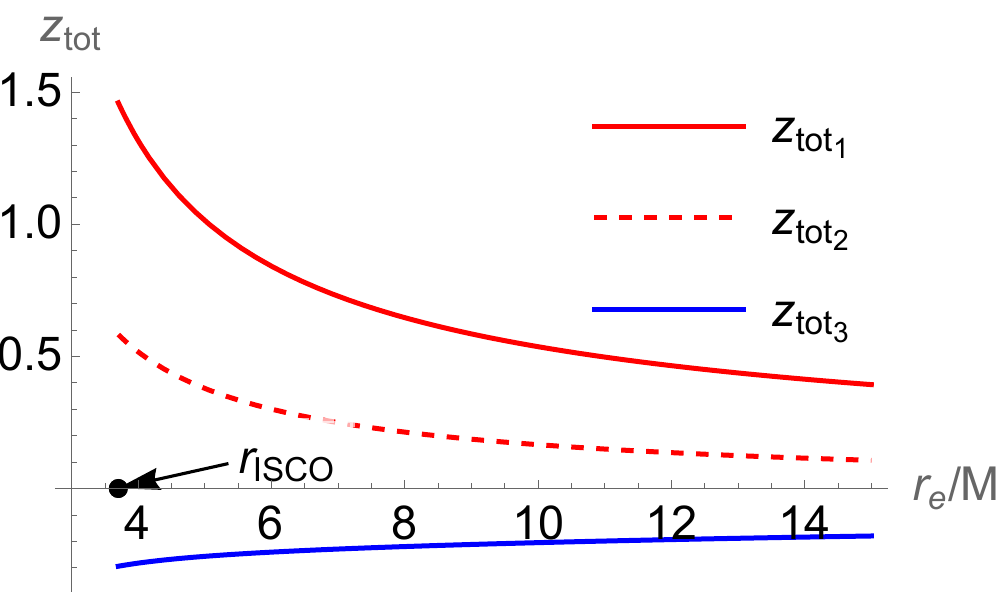}}
\subfigure[]  {\label{z0}
\includegraphics[width=7cm]{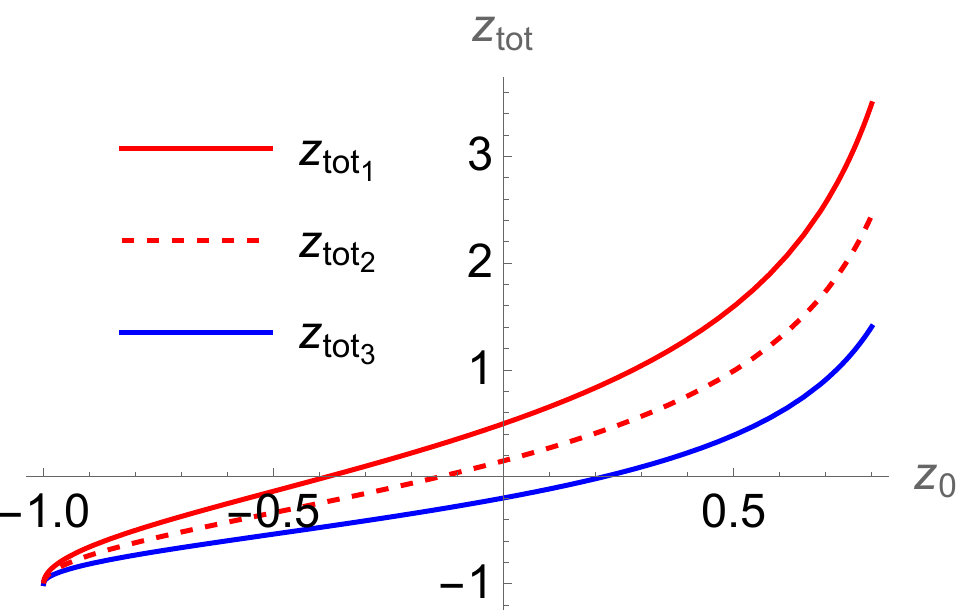}}
\end{center}
\caption{The top panel plots the frequency shifts versus the quantum parameter $A_{\lambda}$ with other {dimensionless parameters set to $\tilde{M}=1/7$} and $z_0=0$. The middle panel shows the frequency shifts versus {the dimensionless orbiting radius $r_{\text{e}}/M$} with $z_0=0$ and $A_{\lambda}=1/3$. The bottom panel shows the frequency shifts versus the peculiar redshift $z_0$ with {$\tilde{M}=1/11$} and $A_{\lambda}=1/3$. {It should be noted that the parameter $A_{\lambda}$ can be the order of unity only for black holes with Planck mass scale, and is extremely small for astrophysical black holes.} Besides, all parameters are set to ensure $E^2>0$, $L^2>0$ and the stable condition (\ref{Veffdd}) and $r_{\text{e}}$ is larger than the radius of the innermost stable circular orbit (ISCO).
}
\label{RST}
\end{figure}

In the following, we will give a brief analysis about the influences of the quantum effects and peculiar velocity on the frequency shifts of photons. Figure \ref{A} shows that both of the redshift $z_{\text{tot}_1}$ and the gravitational redshift $z_{\text{tot}_3}$ decrease with $A_{\lambda}$. Besides, the blueshift also decreases with $A_{\lambda}$ although the value of $z_{\text{tot}_2}$ increases with it, since $\omega_{\text{d}}$ is in the denominator and the smaller $\omega_{\text{d}}$ means the larger $z_{\text{tot}_2}$ but the smaller blueshift. In a word, the quantum effects reduce the frequency shifts of photons. The possible reason is that the quantum effects usually result in a weakening of the gravitational force \cite{Caldwell031301}. What's more, Fig.~\ref{re} shows that the frequency shifts of the photons also decrease with the orbital radius $r_\text{e}$ for fixed $A_{\lambda}$ because the larger $r_\text{e}$ is, the weaker the gravitational field becomes. Finally, Fig.~\ref{z0} shows the frequency shifts versus the peculiar redshift or peculiar velocity. Obviously, the redshift $z_{\text{tot}_1}$ and gravitational redshift $z_{\text{tot}_3}$ increase with the radial receding peculiar velocity (corresponding to positive $z_0$) of the black hole with respect to the detector while the blueshift decreases with it although the value of $z_{\text{tot}_2}$ increases. Besides, there exists a zero point for the blueshift since the kinematic effects is counteracted by the peculiar redshift at a particular receding velocity and similar zero point also respectively appears for the redshift and gravitational redshift at a particular approaching velocity (corresponding to negative $z_0$).

\section{The presence of plasma}~\label{RBp}

In this section, we consider a more practical scenario, i.e., the black hole surrounded by a nonmagnetized, cold plasma, and investigate the effects of the plasma on the redshift/blueshift. The plasma is assumed as a static, inhomogeneous medium and the Hamiltonian for the photon in this plasma is given by \cite{Rogers17,Perlick104031}
\begin{eqnarray}
H=\frac{1}{2}\big[g^{\mu\nu}k_{\mu}k_{\nu}+\omega_{\text{p}}(r)^2\big],
\end{eqnarray}
where $\omega_{\text{p}}$ stands for the electron plasma frequency. By introducing the refractive index of the plasma medium
\begin{eqnarray}
n^2=1-\frac{\omega_{\text{p}}^2}{\omega^2},
\end{eqnarray}
where $\omega$ stands for the frequency of photons measured by a static observer with respect to the plasma, the Hamiltonian can be rewritten as
\begin{eqnarray}~\label{effH}
H=\frac{1}{2}\big[g^{\mu\nu}k_{\mu}k_{\nu}-(n^2-1)(k_{\mu}V^{\mu})^2\big] =\frac{1}{2}g^{\mu\nu}_{\text{eff}}k_{\mu}k_{\nu},~~~
\end{eqnarray}
where  $V^{\mu}$ denotes the four velocity of the plasma, and $g^{\mu\nu}_{\text{eff}}\equiv g^{\mu\nu}-(n^2-1)V^{\mu}V^{\nu}$. Equation (\ref{effH}) shows that photons travel along the geodesic in the effective metric $g^{\mu\nu}_{\text{eff}}$. Thus, for the case of a circular and equatorial orbiting massive particle, Eq.~(\ref{zcomplete}) can still be formally expressed as
\begin{eqnarray}~\label{zplasmasimplify}
1+z=\frac{(E_{\gamma}U^t-L_{\gamma}U^{\phi})|_{\text{e}}}{(E_{\gamma}U^t-L_{\gamma}U^{\phi})|_{\text{d}}}=\frac{U^t_{\text{e}}-\hat{b}_{\text{e}}U^{\phi}_{\text{e}}}{U^t_{\text{d}}-\hat{b}_{\text{d}}U^{\phi}_{\text{d}}},
\end{eqnarray}
with $\hat{b}\equiv \frac{L_{\gamma}}{E_{\gamma}}$. Although the definition is the same as the case of the absence of plasma, the impact parameter now incorporates the effects of plasma as explained in the following. 

Considering the following plasma frequency with a radial power-law number density \cite{Rogers17,Perlick104031}
\begin{eqnarray}
\omega_{\text{p}}(r)^2=\frac{4\pi e^2}{m}N(r), \quad N(r)=\frac{N_0}{r^h},
\end{eqnarray}
where $N_0$ is a constant, and $e$ and $m$ respectively represent the electron charge and mass, the refractive index of the plasma medium can be expressed as
\begin{eqnarray}
n^2=1-A(r)\frac{k}{r^h},
\end{eqnarray}
with $k\equiv\frac{4\pi e^2}{m \omega_0^2}N_0$ and $\omega_0$ the frequency of the photon measured by a detector at infinity. For simplicity, we focus on the case of $h=1$. Within this effective geometry, the energy and angular momentum for photons moving along the geodesics can be obtained as $E_{\gamma}=\frac{A}{n^2}k^t$ and $L_{\gamma}=C k^{\phi}$. Then, the equation of motion for photons at the equatorial plane can be given by
\begin{eqnarray}
-\frac{n^2}{A}E_{\gamma}^2+B(k^r)^2+\frac{L_{\gamma}^2}{C}=0.
\end{eqnarray}

Following the same procedure in the last section, the impact parameter can be calculated as
\begin{eqnarray}~\label{bhat}
\hat{b}=-\frac{\sqrt{C}n\sin\phi}{\sqrt{A(\sin^2\phi+B\cos^2\phi)}}.
\end{eqnarray}
Obviously, the impact parameter incorporates the refractive index $n$ of the plasma medium.
Inserting the impact parameter into Eq.~(\ref{zplasmasimplify}), the total frequency shift with the presence of plasma can be expressed as
\begin{eqnarray}
\hat{z}_{\text{tot}}&=&\sqrt{\frac{1+z_0}{1-z_0}}\bigg(\sqrt{\frac{C'(r_\text{e})}{A(r_\text{e}) C'(r_\text{e})-C(r_\text{e}) A'(r_\text{e})}} \nonumber\\
&+&\frac{n(r_{\text{e}})\sin\phi \sqrt{\frac{C(r_\text{e})A'(r_\text{e})}{A(r_\text{e}) C'(r_\text{e})-C(r_\text{e}) A'(r_\text{e})}}}{\sqrt{A(r_\text{e})(\sin^2\phi+B(r_\text{e})\cos^2\phi)}}\bigg)-1,
\end{eqnarray}
where the static observer is also assumed to be located at infinity with the four-velocity $U^{\mu}|_{\text{d}}=(1,0,0,0)$.

Similarly, one can also define three variables
\begin{subequations}~\label{RSThd}
\begin{eqnarray}
\hat{R}&\equiv& 1+\hat{z}_{\text{tot}_1}, \\
\hat{S}&\equiv& 1+\hat{z}_{\text{tot}_2}, \\
\hat{T}&\equiv& \left(1+\hat{z}_{\text{tot}_3}\right)^2,
\end{eqnarray}
\end{subequations}
where $\hat{z}_{\text{tot}_1}\equiv\hat{z}_{\text{tot}}|_{\phi\rightarrow \pi/2}$, $\hat{z}_{\text{tot}_2}\equiv\hat{z}_{\text{tot}}|_{\phi\rightarrow -\pi/2}$, and $\hat{z}_{\text{tot}_3}\equiv\hat{z}_{\text{tot}}|_{\phi\rightarrow 0}$. Then, two elegant closed formulas for the mass of the black hole and the quantum parameter can be derived from these variables by inverting the relations Eq.~(\ref{RSThd}):
\begin{widetext}
\begin{eqnarray}
M&=&\frac{n r_\text{e} \left(\hat{T}  n^2 (\hat{T} -\zeta )+(\zeta +\hat{T} ) (\hat{R} \hat{S}-\hat{T} )\right)}{\hat{T}  \left(\hat{R} \hat{S}-\hat{T}  \left(1-n^2\right)\right)}\sqrt{\frac{3\zeta  \hat{T} }{\hat{T}  n^2 (7 \zeta -4 \hat{T} )-4 (2 \zeta +\hat{T} ) (\hat{R} \hat{S}-\hat{T} )}}, \\
A_{\lambda}&=&\frac{\hat{T}  \left(\hat{R} \hat{S}-\hat{T}  \left(1-n^2\right)\right)^2 \left(\hat{T}  n^2 (\hat{T} -\zeta )+(2 \zeta +\hat{T} ) (\hat{R} \hat{S}-\hat{T} )\right)}{6 \zeta  \left(\hat{T}  n^3 (\hat{T} -\zeta )+n (\zeta +\hat{T} ) (\hat{R} \hat{S}-\hat{T} )\right)^2}.
\end{eqnarray}
\end{widetext}

\begin{figure}[htb]
\begin{center}
\subfigure  {\label{z1p}
\includegraphics[width=7cm]{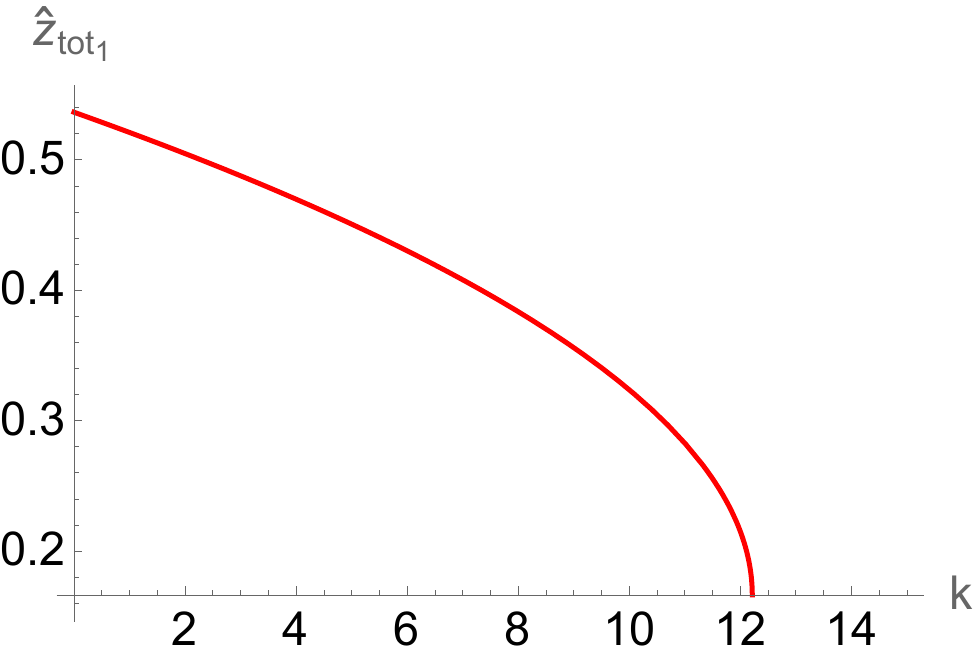}}
\subfigure  {\label{z2p}
\includegraphics[width=7cm]{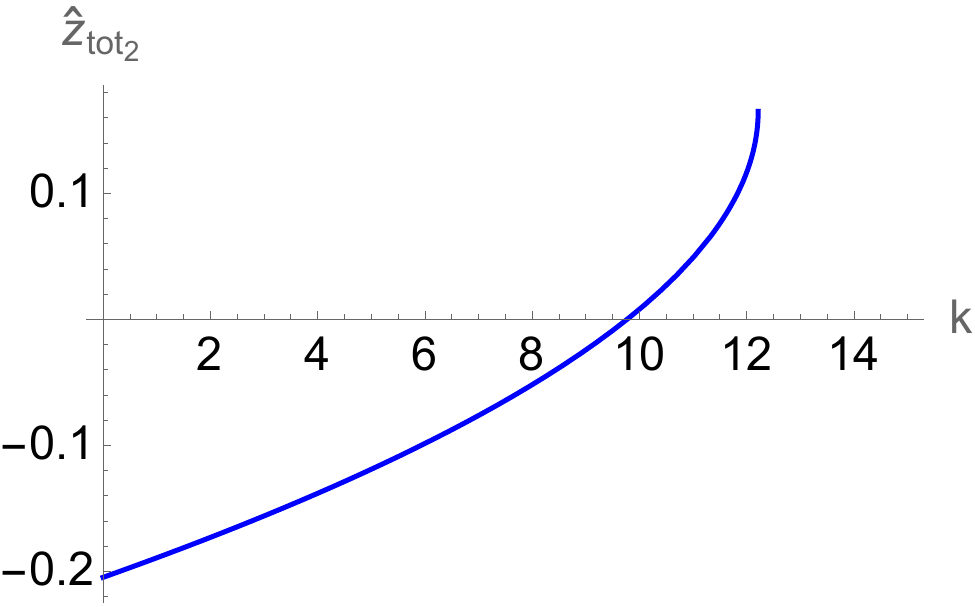}}
\end{center}
\caption{The redshift (top panel) and blueshift (bottom panel) versus the parameter $k$ with vanishing peculiar redshift $z_0=0$. {The other dimensionless parameters are set to $\tilde{M}=1/10$} and $A_{\lambda}=1/3$. All parameters satisfy $E^2>0$, $L^2>0$ and the stable condition (\ref{Veffdd}).}
\label{RpSp}
\end{figure}

In the following, we make a brief analysis about the effects of the plasma on the redshift/blueshift. To make the analysis clearly, we assume fixed $\omega_0$ or $\omega_{\text{d}}$, which is the natural frequency of the photon measured by a static observer located at infinity, and investigate the effects of the plasma on the frequency shifts through the emitting frequency $\omega_{\text{e}}$. Figure \ref{RpSp} shows that both of the redshift and blueshift decrease with $k$, and there is a bound for the range of $k$ since the photons cannot propagate in the region with $\omega_0<\omega_{\text{p}}(r)\sqrt{A(r)}$. Besides, it should be noted that the plasma has no effects on $\hat{z}_{\text{tot}_3}$. The reason is that the impact parameter vanishes when the photon propagate radially along the line of sight of the observer, and Eq.~(\ref{zplasmasimplify}) shows that the influence of the plasma vanishes when $\hat{b}\rightarrow 0$.

\section{Conclusion} ~\label{con}

In this paper, we investigated the frequency shifts of photons emitted by massive particles revolving around a {particular} polymerized black hole {introduced in Refs.~\cite{Bodendorfer136390,Bodendorfer095002}}, which is characterized by two parameters, the black hole mass $M$ and the non-negative parameter $A_{\lambda}$ originated from quantum corrections. It is found that these two parameters can be completely and independently given by few direct astrophysical observables related to the frequency shifts of the photons. Then, we gave a brief analysis about the influences of the quantum effects on the frequency shifts. Besides, since plasma will change the behavior of the photon and most real black holes are surrounded by plasma, we studied the effects of the plasma on the frequency shifts of photons. 

To be specific, for the case of a circular and equatorial orbiting massive particle, we computed the total frequency shift of the photons emitted by the massive particle. {By expanding the total frequency shift for large orbits of the emitter, low peculiar redshifts, and small azimuthal angle, we find that the leading contributions to the frequency shift originate from the kinematic shift  and peculiar shift, and the quantum effects make a negligible contribution to it since $A_{\lambda}$ first appears in the ninth term of the series expansion and the quantum parameter is largely suppressed by the masses of astrophysical black holes.} 

The total frequency shift $z_{\text{tot}}$ includes three special cases, which respectively correspond to the max redshift $z_{\text{tot}_1}$ for $\phi=\frac{\pi}{2}$, the max blueshift $z_{\text{tot}_2}$ for $\phi=-\frac{\pi}{2}$, and the gravitational redshift $z_{\text{tot}_3}$ for $\phi=0$. By solving an inverse problem, we obtained two closed analytic expressions for the black hole mass and quantum parameter in terms of the redshift $z_{\text{tot}_1}$, blueshift $z_{\text{tot}_2}$ and gravitational redshift $z_{\text{tot}_3}$, as well as the orbital radius $r_{\text{e}}$ of the massive particle and the peculiar redshift $z_0$ which codifies the peculiar motion of the black hole. Namely, the mass and the quantum parameter {describing this particular polymerized black hole} can be completely determined by few direct observables. Besides, we found that all frequency shifts of photons decrease with the quantum parameter $A_{\lambda}$ since the quantum effects weaken the strength of the gravitational field. The frequency shifts also decrease with $r_{\text{e}}$ since the larger orbital radius means a weaker gravitational field. In addition, we investigated the frequency shifts versus the peculiar redshift and found that the redshift and gravitational redshift increase with the receding peculiar velocity of the black hole while the blueshift decreases with it. There exists a zero point for the blueshift where the kinematic effect is counteracted by the peculiar redshift at a particular receding velocity and  similar zero point also respectively appears for the redshift and gravitational redshift at a particular approaching velocity.

Then, we investigated the refractive plasma effects on the frequency shifts of photons. After introducing an index of refraction of the plasma medium, the photons can still travel along the geodesic of an effective metric. Then, we calculated the frequency shifts of the photons following the same way as in the case of the absence of the plasma, and found that the black hole mass and quantum parameter also can be analytically solved in terms of the redshift $\hat{z}_{\text{tot}_1}$, blueshift $\hat{z}_{\text{tot}_2}$ and gravitational redshift $\hat{z}_{\text{tot}_3}$. Besides, we analyzed the dispersion effects of the plasma and found that the frequency shifts always decrease with the plasma density.

Finally, we stress that this general relativistic formalism has been applied to a lot of Schwarzschild black holes hosted at the core of different active galactic nuclei, such as, NGC 4258 \cite{NucamendiL14}, TXS-2226-184 \cite{VillalobosL9}, J1346+5228 \cite{Villalobos06486}, and so on, where the authors estimate the mass-to-distance ratio of these black holes with a Bayesian fitting method by using the large data set of the frequency shifts of photons emitted by water masers orbiting them. In our forthcoming work, we will estimate the mass of the polymerized black hole hosted at different active galactic nuclei and evaluate the quantum effects with the general relativistic formalism.

\acknowledgments{
This work was supported by
the National Natural Science Foundation of China (Grants Nos. 11975072, 11875102 and 11835009),
the National SKA Program of China (Grants Nos. 2022SKA0110200 and 2022SKA0110203),
the China Postdoctoral Science Foundation (Grant No. 2021M700729), 
and Shaanxi Provincial Education Department (Grant No. 21JK0556).
}

\end{document}